# Navigating Cybersecurity Training: A Comprehensive Review

Saif Al-Dean Qawasmeh, Ali Abdullah S. AlQahtani, *Member, IEEE*, Muhammad Khurram Khan, *Senior Member, IEEE*

*Abstract*—In the dynamic realm of cybersecurity, awareness training is crucial for strengthening defenses against cyber threats. This survey examines a spectrum of cybersecurity awareness training methods, analyzing traditional, technology-based, and innovative strategies. It evaluates the principles, efficacy, and constraints of each method, presenting a comparative analysis that highlights their pros and cons. The study also investigates emerging trends like artificial intelligence and extended reality, discussing their prospective influence on the future of cybersecurity training. Additionally, it addresses implementation challenges and proposes solutions, drawing on insights from real-world case studies. The goal is to bolster the understanding of cybersecurity awareness training's current landscape, offering valuable perspectives for both practitioners and scholars.

## I. INTRODUCTION

As of early 2023, the global internet user base has soared to approximately 5.44 billion, accounting for 68% of the global population, up from 3.977 billion in 2018 [1]. This surge encompasses diverse sectors such as education, commerce, and social media, the latter boasting over 3 billion users [2].

The COVID-19 pandemic has further catalyzed the digital transition, with a notable pivot to online learning and remote working [3], and a boom in e-commerce [4]. However, this increased reliance on digital platforms has escalated the threat of cybercrime, with costs exceeding 945 billion USD in 2020, alongside a substantial investment in cyber defenses [5].

The pandemic-induced shift to remote work has opened new avenues for cybercriminals, intensifying the frequency and sophistication of attacks, especially through social engineering and phishing [6]–[10]. Despite technological strides in Artificial Intelligence (AI) and Machine Learning (ML) for combating such threats [11], [12], the human factor remains a critical vulnerability [13], [14]. Empirical evidence suggests that enhancing cybersecurity awareness significantly bolsters security behaviors and policy compliance [15].

Organizations employ a variety of training methods to educate employees about cybersecurity threats, ranging from tradi-

S. Qawasmeh and A. AlQahtani are North Carolina A&T State University, Greensboro, NC, USA, 27411.
E-mails: Qawasmeh.saif1@gmail.com, AlQahtani.aasa@gmail.com
M.K. Khan is with the Center of Excellence in Information Assurance (CoEIA), King Saud University, Saudi Arabia.
E-mail: mkhurram@ksu.edu.sa

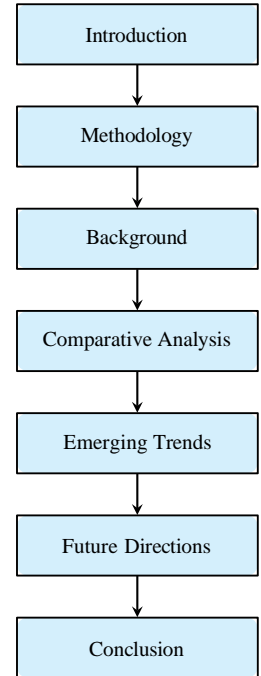

Fig. 1: Paper Structure

tional classroom settings to innovative Virtual Reality (VR) and Augmented Reality (AR) simulations [11], [14], [16]–[39]. Despite the variety of cybersecurity awareness training methods available, comprehensive studies discussing these methods and their effectiveness are limited. This paper aims to address this gap by providing a comprehensive survey of state-of-the-art methods in cybersecurity awareness training. We conduct a diverse research methods, see Section II, to study the effectiveness of these methods and provide an overview of their advantages and disadvantages. Our goal is to guide organizations in selecting the most suitable training methods for their specific needs, ultimately enhancing their defense against cyber-attacks.

This paper addresses key questions in the field of cybersecurity awareness training:

1) What methodologies are currently used, and how have they developed over time?
2) How do these methodologies compare in terms of strengths, weaknesses, and effectiveness?
3) What emerging trends are shaping the future of cybersecurity training, and what implementation challenges exist?

4) What future developments can we expect in cybersecurity awareness training?

Our survey distinguishes itself from recent literature [13], [20], [40]–[48] in several ways:

1) Unlike [13], [20], [40]–[48], it provides a detailed examination of current cybersecurity training methodologies, setting it apart from previous surveys.
2) Our analysis integrates individual method characteristics with a comparative evaluation, offering deeper insights than [41], [45].
3) Unlike recently published surveys [13], [20], [40], [41], [43]–[48], we explore recent trends like AI, VR, and AR in training, addressing challenges in their implementation, which is not extensively covered in [42].
4) The paper discusses the challenges and future directions for cybersecurity awareness training, unlike [40], [42], [44], [48].
5) We assess mobile and web applications as standalone methods, providing a unique perspective compared to [20], [45].
6) Our categorization of training methods into sub-categories allows for a more nuanced assessment, differing from the approach in [13], [43], [46], [47]; Table I shows a comparison between this paper and other surveys.

The remainder of this paper is structured as follows: Section II describes the methodologies employed to address our research questions. Section III presents the training methods and reviews the evolution and current trends in cybersecurity awareness training. Section IV provides a comparative analysis of these methods, assessing their effectiveness and engagement metrics. Section V looks at emerging trends in cybersecurity training and their potential impact. Section VI discusses challenges in training implementation and proposes solutions. Section VII concludes the paper with a summary of our findings. Figure 1 illustrates the overall structure visually.

## II. METHODOLOGY

Our approach to addressing the research questions combines a systematic literature review, extensive online searches, and personal observations. We analyzed primary and secondary data from 63 studies, focusing on eight distinct cybersecurity training delivery methods. The literature review was divided into a general search, yielding 255 papers from an initial set of 476, and a specific search based on delivery methods, resulting in 67 papers suitable for review, see Table II. Keywords such as "Cybersecurity awareness methods," "Information security awareness," and "Electronic information security" guided our search, with additional specific keywords for each training method.

The online search extended across databases including Google Scholar, ScienceDirect, Scopus, Springer, JSTOR, ACM Digital Library, IEEE Xplore, Emerald, and John Wiley online libraries. This ensured a comprehensive collection of relevant literature.

Our observations draw upon academic experience and insights from the literature, categorizing methodologies by effectiveness in raising cybersecurity awareness and proposing evaluation criteria for these methods. A word cloud visualizes the most common keywords and titles from our search (Figure 2). Despite a thorough approach, our study acknowledges limitations such as the focus on English-language sources and the primary reliance on academic and industry literature, which may overlook non-English and non-mainstream initiatives.

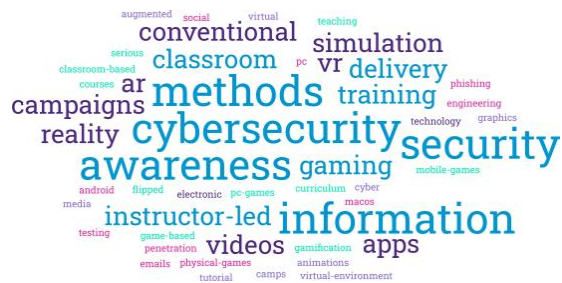

Fig. 2: Word cloud generated from the most common keywords and titles

## III. BACKGROUND

This section presents the current methodologies in cybersecurity awareness training, and how have they evolved over time. Also, we present a systematic review of the prevailing methodologies in cybersecurity awareness training, categorized into three groups: traditional, technology-based, and innovative training methods.

### A. Traditional Training Methods

Traditional training methods include tools that do not, usually, rely on advanced technologies and are used to explain cybersecurity concepts. Conventional methods are centered around implementing IS training for large populations, such as paper-based and electronic flyers, lectures, courses, educational videos, and cyber campaigns; Figure 3 illustrates some examples of traditional training methods used to increase cybersecurity awareness.

*1) Passive Awareness Methods:* In the realm of cybersecurity awareness, the transition from traditional passive learning methods like posters and flyers to digital mediums marks a significant evolution. Originally, approaches like those employed by The European Union Agency for Cybersecurity (ENISA) and Minnesota State Colleges and Universities relied on static posters to educate about password usage, as shown in Figure 4 [119]–[122]. These methods, while effective in their era, were limited by their static nature and often faded into the background in bustling environments.

The digital transformation has ushered in a new era of e-posters, e-articles, and emails [123], [124]. This shift not only expands the audience reach but also introduces dynamic content updates, crucial in the fast-evolving cybersecurity domain. Digital tools contrast starkly with their physical counterparts,

TABLE I: Comparison with other surveys, where ✓: cover the topic, ✗: does not cover the topic

| Method<br>Study | Conventional | Classroom | Gamification | Simulation | Applications | Videos | Campaigns | VR/AR |
|---|---|---|---|---|---|---|---|---|
| [40] | ✓ | ✓ | ✓ | ✓ | ✗ | ✓ | ✗ | ✗ |
| [13] | ✓ | ✓ | ✓ | ✓ | ✗ | ✓ | ✗ | ✗ |
| [41] | ✗ | ✗ | ✓ | ✓ | ✗ | ✗ | ✗ | ✗ |
| [42] | ✗ | ✗ | ✓ | ✓ | ✗ | ✗ | ✓ | ✗ |
| [20] | ✓ | ✓ | ✓ | ✗ | ✓ | ✓ | ✗ | ✗ |
| [43] | ✓ | ✓ | ✓ | ✓ | ✗ | ✓ | ✓ | ✗ |
| [44] | ✗ | ✗ | ✓ | ✗ | ✗ | ✓ | ✗ | ✗ |
| [45] | ✓ | ✓ | ✓ | ✓ | ✗ | ✓ | ✓ | ✗ |
| [46] | ✓ | ✓ | ✓ | ✗ | ✗ | ✓ | ✗ | ✗ |
| [47] | ✓ | ✓ | ✓ | ✓ | ✗ | ✓ | ✗ | ✗ |
| [48] | ✗ | ✓ | ✓ | ✗ | ✓ | ✓ | ✓ | ✗ |
| Ours | ✓ | ✓ | ✓ | ✓ | ✓ | ✓ | ✓ | ✓ |

TABLE II: Surveys Summary Categorized by Cybersecurity Training Delivery Methods

| Delivery Method | Studies | Number of Studies |
|---|---|---|
| Conventional | [49] [50] [51] [52] [53] [54] [55] | 7 |
| Classroom | [24] [56] [57] [58] [59] [60] [61] [62] [63] [64] [65] | 11 |
| Gamification | [66] [67] [68] [69] [70] [71] [72] [73] [74] [75] | 10 |
| Simulation | [76] [77] [78] [79] [80] [81] [82] [83] [84] [85] [86] | 11 |
| Applications | [87] [88] [89] [90] [91] [35] [92] [93] [94] [95] | 10 |
| Videos | [22] [21] [96] [97] [98] [99] [100] [101] [102] [103] | 10 |
| Campaigns | [104] [105] [106] [107] [108] [109] [110] [111] [112] [113] | 10 |
| VR/AR | [114] [115] [116] [117] | 4 |

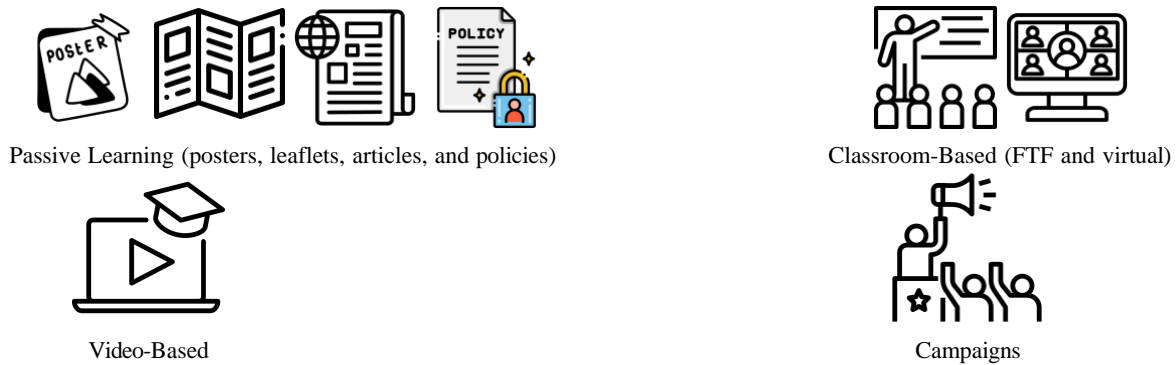

Fig. 3: Traditional Learning Methods; Icons From [118]

offering scalable and timely dissemination of vital security messages.

This transition from physical to digital is a strategic response to the ever-changing landscape of cyber threats. In today's digital age, cybersecurity awareness methods must adapt with the same agility as the threats they address. Digital media ensures that the intended messages not only reach a wider audience but also engage them effectively, creating a more informed and vigilant community.

While digital methods are gaining prominence, traditional methods like posters and articles continue to play a key role in cybersecurity awareness, often complementing other security strategies [125]. These approaches, along with legal frameworks defining cyber-attack laws, provide a multi-faceted approach to awareness [126], [127].

Research, such as that by Khan et al. [49], has explored the impact of various Information Security Awareness (ISA) tools, proposing metrics for workplace effectiveness. In the banking sector, a blend of conventional, instructor-based, and digital methods has been identified, with visually engaging materials like posters proving particularly effective [50]. Universities also employ diverse methods, including Information Security Awareness Training (ISAT) programs featuring regular updates and tutorials [51].

Although physical posters can lead to information saturation, focused studies have shown that well-designed posters can still effectively reinforce cybersecurity concepts [52]. On the digital front, interactive elements in e-articles and e-posters guide users to additional resources, complemented by regular newsletters and alert emails [53]. These methods' effectiveness is further supported by research applying the Protection Motivation Theory (PMT) [54].

Interactive workshops have also emerged as a vital component in cybersecurity awareness. Albrechtsen and Hovden's work [55] underscores the value of small-group workshops in enhancing cyber-awareness within organizational settings. This multifaceted approach, blending both traditional and digital methods, ensures a comprehensive and adaptive strategy in promoting cybersecurity awareness.

*2) Classroom-Based:* In the field of cybersecurity education, we've witnessed a significant transformation from traditional classroom-based training to more versatile and accessible

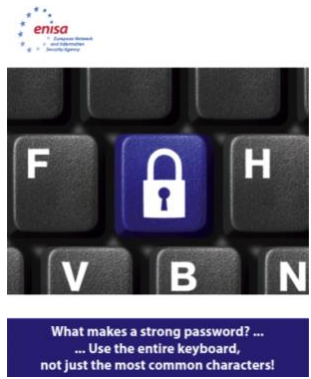
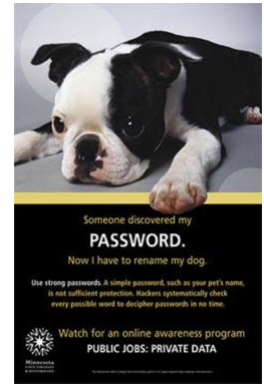

ENISA Password Awareness Poster    Minnesota State Colleges and Universities Password Awareness Poster

Fig. 4: Cybersecurity Awareness Poster Examples [122]

forms. Initially, cybersecurity education was predominantly delivered through face-to-face instruction, a method effective yet limited by logistical constraints [96].

The emergence of virtual learning environments, exemplified by platforms like Udemy and Lynda, has revolutionized this landscape. These platforms enable both synchronous and asynchronous learning, overcoming the limitations of time and location [128]. This transition to online education has retained the effectiveness of traditional methods while introducing the benefits of self-paced learning, allowing learners to assimilate information at their own pace [21], [22].

An innovative blend of online and traditional methods is seen in the flipped classroom model. This approach combines the autonomy of online learning with the interactive experience of physical labs, aligning with HyFlex pedagogical principles that emphasize student choice, learning outcome equivalency, content reusability, and tool accessibility [129], [130].

The evolution of cybersecurity training reflects a broader pedagogical shift from teacher-centered to learner-centered models. Modern cybersecurity curricula are designed to cater to a range of topics, from foundational principles to advanced threat defenses, accommodating diverse learning styles and lifestyles [60], [131]–[133].

This transformation underscores the need for adaptable teaching strategies that resonate with a generation of learners seeking flexibility, engagement, and practicality in their education. Classroom-based cybersecurity training today encompasses a spectrum of methods, from interactive in-person classes to self-paced online courses [134], [135]. Studies like Ali's, which explored lecture-based training including quizzes on cyber threats, demonstrate improved student understanding of cybersecurity risks [24]. Carella et al.'s research validates the effectiveness of classroom training in enhancing awareness and reducing phishing susceptibility [56].

Instructor-led sessions remain crucial, with comparative studies highlighting the benefits of gamified learning over traditional lectures [57]. Kim et al.'s work comparing Instructor-Based Training (IBT) to Computer-Based Training (CBT) underscores the value of hands-on experiences in cybersecurity education [58].

Academic integration of cybersecurity education is increasingly advocated. Khader's Cyber-security Awareness Framework for Academia (CAFA) suggests a gamified approach [60]. Churi and Rao, as well as Luburic et al., propose combining traditional teaching with interactive elements like lab exercises, emphasizing the flipped classroom model's effectiveness [61], [62].

Innovative methods are also being explored, such as Cai and Arney's Top-down & Case-driven (TDCD) approach, integrating theory with practical labs [63], and Konstantinou's course combining lectures with hands-on labs [64]. Tran et al. introduced a game-based, self-regulated course, showcasing gamification's potential in cybersecurity education [65].

This paradigm shift in cybersecurity training, embracing both traditional and innovative methods, reflects a commitment to providing diverse and effective educational experiences in response to the evolving landscape of cyber threats and learner preferences.

*3) Video-Based:* The transformation of video-based cybersecurity education mirrors the broader shift towards more dynamic and engaging pedagogical methods. Initially, video learning in cybersecurity was largely limited to straightforward lectures. While these were informative, they didn't fully utilize the medium's potential for interaction and engagement. Now, the landscape includes a much more diverse array of video formats, such as interactive and 360-degree experiences, which significantly enhance learner engagement [136], [137].

This evolution from passive viewing to active engagement in video-based learning is a direct response to the need for more dynamic educational methods. These methods cater to various learning styles and address the challenge of retaining learner attention, a task increasingly difficult in today's digitally saturated environment.

Research has been conducted to assess the efficacy of different video styles. For example, Kohler et al. [138] evaluated the impact of traditional, interview-based, and animated videos on phishing awareness. Despite the variations in format, each style proved effective in enhancing awareness and aiding memory retention. However, He et al. [98] found that some learners still prefer text-based learning, especially when considering self-efficacy and knowledge retention, indicating the importance of tailoring video-based learning to individual needs.

The Open Cyber University's use of pre-recorded, talking-head style lectures supplemented with interactive materials highlights the potential of video-based learning to improve cognitive processing skills [139]. Additionally, platforms recognized by Expert Insights have integrated video training into their cybersecurity awareness programs, showcasing the medium's versatility; see Figure 5.

Further innovations, such as 360-degree and interactive videos, have enriched the video-based learning paradigm. These formats provide immersive experiences that engage learners and promote better knowledge retention through interactive elements like quizzes and games [140].

The development of video-based cybersecurity training underscores the shift in educational strategies towards more engaging and interactive content delivery. In a world overflowing with digital content, maintaining learner engagement through innovative video techniques is essential. These methods use dynamic visuals to enhance cybersecurity education [141]. Alkhazi et al. [22] demonstrated that a combination of lectures, videos, and text materials significantly improves cybersecurity attitudes among employees compared to lectures alone. Similarly, Sutter et al. [21] found that a mix of videos, text, and quizzes was effective in anti-phishing training.

Comparative research has shown that video-based training can be as effective as other methods for teaching about cybersecurity threats like password security and malware [97]. He et al. [98] revealed that cybersecurity video-based training positively affects self-efficacy. Tutorial videos, such as those developed by Volkamer et al. [99], provide a focused way to educate users on specific threats like phishing.

Studies by Reinheimer et al. [100] and Jones et al. [101] indicate that videos can be a preferred learning medium for certain groups, and when combined with interactive elements, can significantly increase awareness. Shaw et al. [103] concluded that integrating video with other multimedia formats can further enhance the effectiveness of cybersecurity awareness programs, with hypertext being especially impactful.

In summary, the evolution of video-based cybersecurity training from basic lectures to a variety of interactive formats reflects an important shift in educational methods. This evolution caters to the changing needs and preferences of learners, ensuring that cybersecurity education remains effective and relevant in the digital age.

*4) Cyber Campaigns:* The evolution of cyber campaigns reflects the increasingly complex landscape of cybersecurity threats and the need for more sophisticated awareness strategies. Initially, these campaigns were simple, often limited to emails or updates reminding users of best practices in cybersecurity. As cyber threats have advanced, so too have the efforts to combat them. A prime example is the "STOP. THINK. CONNECT" campaign by the Anti-Phishing Working Group (APWG) and National Cyber Security Alliance (NCSA), which offers a variety of resources like videos and tip sheets to elevate user awareness [37].

Cyber Awareness Month, initiated in 2004, epitomizes the dynamic nature of these campaigns. It has evolved from offering basic advice, such as updating antivirus software, to addressing complex countermeasures like two-factor authentication and anti-phishing techniques. This progression mirrors the increasing sophistication of cyber threats.

The integration of Generative AI into cybersecurity campaigns is a response to the ongoing advancements in technology. Employing AI in future campaigns is a strategic necessity to ensure that awareness initiatives remain relevant and effective in the face of evolving cyber tactics.

Table III showcases various international cyber campaigns, outlining their objectives and strategies. These range from providing online safety tips and enhancing email security to promoting good cyber hygiene practices and educating African citizens on cybersecurity concepts. Each campaign is tailored to its specific demographic, reflecting the global reach and adaptability of these initiatives [37], [142]–[145].

The shift from straightforward information dissemination to complex, interactive operations in cyber campaigns is a direct response to the growing sophistication of cyber threats. This transformation underscores the ongoing need for comprehensive and adaptive cybersecurity awareness strategies.

Eminagaoglu et al.'s campaign for Turkish employees significantly reduced weak password use, illustrating the impact of well-structured campaigns [104]. The "Think Before You Click" initiative at a university brought attention to password security, with notable results among students [105]. Christmann et al. showed the effectiveness of animated campaigns in improving employee password habits [106], while a diverse campaign combined digital and physical activities to bolster cybersecurity defenses [107].

The South African Cyber Security Academic Alliance (SAC-SAA) engaged school children in cybersecurity education, demonstrating the importance of starting awareness at a young age [108]. At the University of Missouri-Columbia (MU), a comprehensive campaign was launched to educate on information security, addressing various threats like phishing [109].

Campaigns like Yeoh et al.'s in Australia focused on reducing phishing incidents and increasing reporting, showing the value of targeted initiatives [110]. The "Let's Go Phishing" campaign successfully decreased phishing susceptibility through a multi-platform approach [111]. Eftimie et al. linked social tendencies with increased phishing risk in their cybersecurity course, highlighting the role of social factors in cybersecurity [112]. Bullee et al. reported a decline in susceptibility to telephone scams following an informative campaign, showcasing the effectiveness of targeted awareness efforts [113].

In summary, the ongoing evolution of cyber campaigns from rudimentary reminders to sophisticated, multi-dimensional operations is a testament to the adaptive nature of cybersecurity

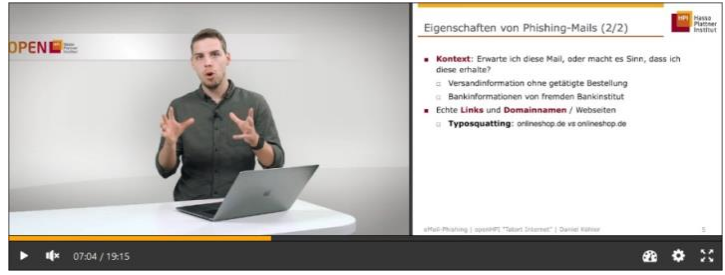
Traditional Video with Presenter

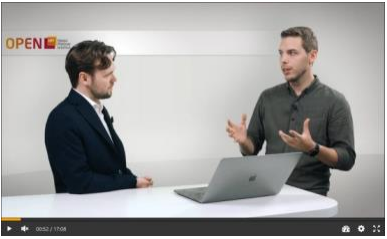
Interview Video

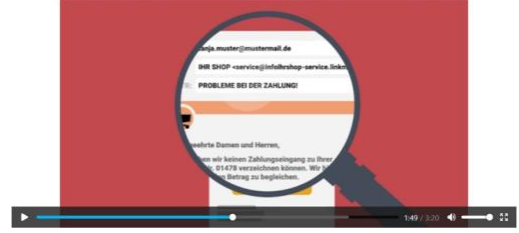
Animated Video

Fig. 5: Screenshots of the cybersecurity educational videos used in [138]

TABLE III: Campaigns examples conducted in different countries

| Campaign | Description | Reference |
|---|---|---|
| Get Safe Online (UK) | The campaign aims to provide tips on how to protect against computers and mobile devices online attacks, security news, and awareness events for the personal and business sectors. | [142] |
| CyberAware (UK) | The campaign conducted by the National Cyber Security Centre (NCSC) aims to provide advice on how to secure emails through the use of strong passwords, data backups, and security updates. | [143] |
| Cyber Streetwise (UK) | A web-based campaign developed by the UK government that aims to enhance online safety and provide users with the knowledge needed to fight against cyber-attacks. The campaign includes several measures to enhance cyber knowledge such as using strong passwords, installing and updating antivirus software, and double-checking the source of online retailers. | [144] |
| STOP.THINK.CONNECT (USA) | A global online safety awareness campaign led by APWG and NCSA to encourage safe online habits, help users adopt good cyber hygiene practices, and engage in the nation's efforts to improve cybersecurity. | [37] |
| Internet Safety Campaign (Africa) | The campaign aims to adopt safe usage of computers and the internet and educate African citizens about the concepts of cybersecurity. | [145] |

awareness strategies in the face of evolving cyber threats and technologies.

### B. Technology-Based Training

In this section, we are shifting from traditional methods to technology-based methods emphasizing the role of technology in enriching learning and empowering learners to effectively confront and address the intricacies of cyber threats.

*1) Simulation-Based:* The cybersecurity training landscape has evolved significantly, transitioning from reactive post-incident measures to proactive, immersive simulations. Initially, cybersecurity training methods were somewhat generic, lacking the necessary sophistication and customization to prepare employees effectively for specific threats they might encounter in their roles [30].

Today, simulation-based training is characterized by its tailored approach. Advanced techniques like spear phishing simulations, personalized to the individual's role within an organization, have become prevalent. This level of customization, achieved by leveraging social engineering skills to craft and distribute phishing emails, has introduced a new degree of realism and personalization to these simulations [31], [32].

Studies by Burns et al. [146] and Jansson et al. [147] have demonstrated the effectiveness of these simulations. They assess employees' awareness levels and provide targeted training to those most at risk, thereby significantly reducing susceptibility to phishing attacks. This training is often conducted between cyber drills to measure its impact on security behavior [148], [149].

The integration of AI into these simulations has been pivotal. AI enables the creation of highly convincing phishing emails that closely mimic legitimate correspondence, extending beyond email generation to include timing of dispatch and analysis of employee responses. This results in a dynamic and responsive training environment [11], [150].

Figures 6 and 7 illustrate the evolution of phishing email templates used in these simulations, showing the progression from less convincing attempts to those nearly indistinguishable from genuine communications [151].

The advancement in simulation-based training reflects a broader trend in cybersecurity education. It marks a shift from general, informative strategies to targeted, interactive simulations that actively engage employees, enhancing their ability to identify and respond to cyber threats. Simulations replicate real-world cyber-attacks to assess user responses and address specific behavioral vulnerabilities [151]–[153].

Workman et al. [76] demonstrated the superiority of simulation-based methods over traditional lectures, incorporating a variety of teaching methods and assessments to gauge their impact on cybersecurity awareness.

Understanding user susceptibility to cyber threats is crucial, particularly regarding phishing. Wright and Marett [77] investigated the influence of computer skills on threat mitigation, while Halevi et al. [78] examined the impact of personal traits on phishing vulnerability, identifying certain demographics as more susceptible.

Other studies assess the effectiveness of training and sanctions in deterring cyber threats. Baillon et al. [154] found that while both cyber-information and simulation training are effective, their combined impact was not synergistic. Kim et al. [82] supported the effectiveness of punitive measures and training, particularly for higher-ranking employees.

The complexity of attacker strategies is further highlighted by simulations like "SpearSim" [83], which evaluate the success of personalized spear-phishing attacks and emphasize the importance of user education.

Research also extends to fake websites and social networking threats, with studies highlighting attacker techniques from urgency cues to website cloning, stressing the necessity for comprehensive training across digital platforms [84]–[86].

In summary, the development of simulation-based training in cybersecurity represents an important shift towards more interactive, personalized, and effective methods of preparing employees to counteract evolving cyber threats.

*2) App-Based:* App-based cybersecurity training has evolved remarkably from being mere information repositories to dynamic, interactive platforms. Initially, these apps provided basic cybersecurity tips and guidelines in a static format, lacking user engagement or progress tracking [34], [35].

Today's cybersecurity applications have become sophisticated tools, engaging users with interactive quizzes, gamification, and real-time updates on cybersecurity threats. Focusing on user experience, they make learning both accessible and engaging [155]. An example is the "STOW" app from Saudi Arabia, offering IS scenarios and quizzes, significantly improving user awareness [156]. Figure 8 depicts two mobile applications, 'LetSecure' and 'STOW', used for this purpose.

This transition to interactive learning is also evident in desktop applications like "HOUSIE," which provides a Q&A format on cybersecurity topics, and mobile apps that offer the latest news on cyber threats [157], [158]. These tools not only impart knowledge but also actively test users' understanding, making learning an engaging process.

The sophistication of these training tools has increased, integrating educational content with interactive features to enhance learning. Applications like SecurityMentor and Webroot's training app offer comprehensive packages, including phishing simulators and quizzes, with enhanced graphics and user-friendly interfaces [159], [160].

The evolution from basic apps to interactive, user-centric platforms reflects technological advancements and a deeper understanding of user engagement in educational technology. This shift indicates a broader trend in cybersecurity awareness methodologies toward more interactive, user-focused learning experiences.

Mobile applications play a crucial role in enhancing information security awareness. Drevin et al. [87] highlighted the impact of mobile apps on user awareness and personality traits. Razaque et al. [88] developed the Web-Based Cyber-security Program (WBCA) to improve understanding of cyber-crimes. Feito-Pin et al. [89] and Aguayo et al. [90] used cross-platform apps and animated narratives, respectively, with quizzes to measure comprehension.

Phishing-focused apps like "Social-Phish" and "Chat-phish" by Bandreddi [92] teach users to distinguish between genuine and fake websites. "Quiz Your Permissions" evaluates broader cybersecurity knowledge, confirming its effectiveness among University of Toledo students.

Antonucci et al. [94] introduced "Pause-and-Think (PAT)", an anti-phishing app that prompts users to critically evaluate email attachments. Their study involving 107 participants showed that warning messages with counters increased phishing awareness.

A quiz-based app studied by [95] proved effective in enhancing users' cybersecurity knowledge, first assessing and then directing them to customized learning content. This progression in app-based cybersecurity training, from simple information delivery to engaging and interactive experiences, demonstrates a significant advance in educational technologies and methodologies in the cybersecurity domain.

### C. Innovative Training Methods

This section focuses on the evolution and impact of innovative training methods in cybersecurity education. As technology reshapes the learning landscape, these advanced techniques play a crucial role in preparing individuals to face the increasingly complex world of cyber threats.

*1) Game-Based:* Game-based learning in cybersecurity has significantly evolved, transitioning from early text-based formats to the current interactive and immersive experiences. Initially, cybersecurity games like "Hacker" were simple, focusing on imparting knowledge through text, with limited user engagement [161].

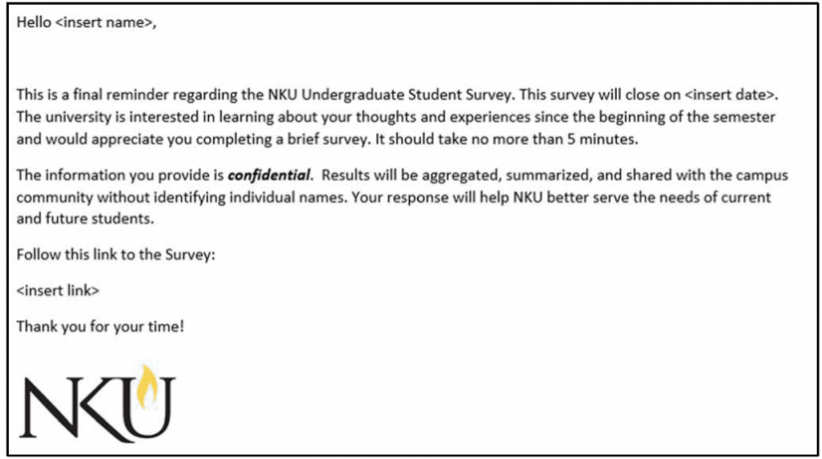

Fig. 6: Phishing email template asking to fill a survey [148]

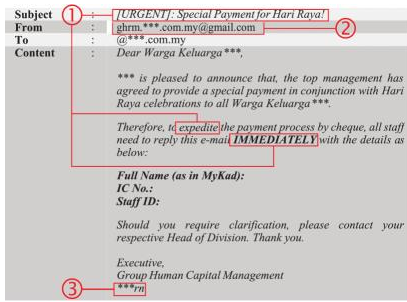

Simulated phishing email before changing content

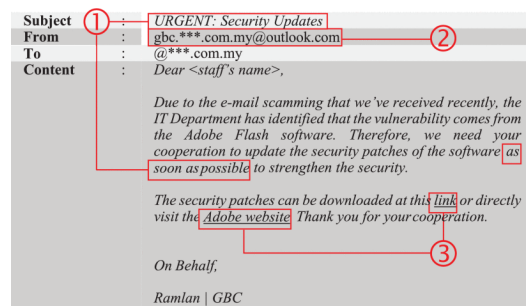

Simulated phishing email after changing content

Fig. 7: Less convincing phishing email (a) VS phishing email that conforms a legitimate email (b) [151]

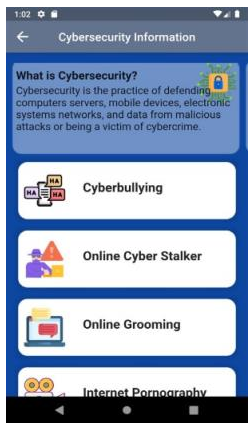

LetSecure App [35]

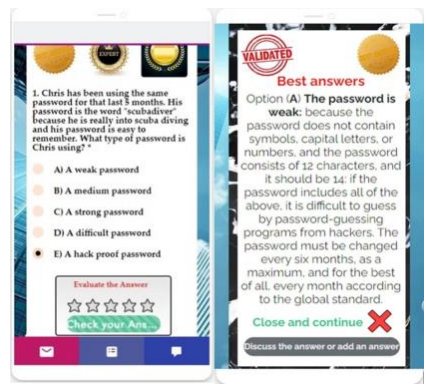

STOW APP [156]

Fig. 8: Awareness mobile apps examples

4In contrast, modern cybersecurity games incorporate rich multimedia content, competitive elements, rewards, and real-time feedback, greatly enhancing learning outcomes. For instance, see figure 9, it effectively tests and improves employees' knowledge across various domains, including password security and phishing awareness [26], [27], [29].

The trend towards "edutainment" in cybersecurity is evident in the development of serious games that combine education with compelling user engagement. These games simulate real-world scenarios, allowing users to learn through experience, a method proven to enhance skill acquisition and retention [28], [74], [162], [163].

Cybersecurity games today are designed to align with users' objectives and address current threats interactively. Some games are even developed in physical formats to enhance knowledge retention, examples are shown in figure 10. Earlier games like "Hacker" provided foundational concepts in a basic 2D format, while contemporary games, such as "Cyber Security Tycoon", offer more complex and graphically rich environments [164]. Figure 11 shows screenshots of both Hacker and Cyber Security Tycoon games.

This shift from text-based to multimedia-rich, interactive gaming in cybersecurity training reflects a broader pedagogical move towards interactive learning. It highlights innovative approaches in enhancing user engagement and knowledge retention in the field.

Mobile and PC-based games have become key tools for cyberawareness. Games like those developed by Arachchilage and Cole teach users to identify safe URLs and emails, and "Soceng Warriors" offers mini-games on cyber attack types [66], [67]. PC-based games, such as the InCTF competition's game and SCIPS, provide detailed graphics and scenarios that teach about cyber attack costs and protection [69], [70]. Games simulating an attacker's perspective, like "SREG" and "CIST", help users understand security vulnerabilities [71], [72]. Jin et al. created a suite of games for high schoolers, covering various cybersecurity topics [73].

In physical spaces, games like "RISKIO" and "CySecEscape 2.0" offer unique learning experiences in cyber warfare and password security, respectively [74], [75].

Overall, the transformation of game-based learning in cybersecurity illustrates an effective blend of technology and pedagogy, advancing user engagement and educational efficacy in this critical field.

*2) Virtual Reality/Augmented Reality:* The integration of Virtual Reality (VR) and Augmented Reality (AR) in cybersecurity training has undergone significant evolution, marking a shift from basic demonstrations to highly interactive and immersive educational experiences. Initially, VR in cybersecurity was limited, providing users with a visual representation of cyber threats but lacking substantial interaction [165].

Today's VR and AR systems offer high-fidelity, hands-on experiences in cybersecurity training. Users can interact with virtual elements realistically, enhancing both learning and retention. A notable example is the "CISE-PROS VR" game, which uses the HTC Vive headset to immerse students in a virtual data center. Here, they perform actions like cable management and hardware replacement, crucial for the Cyber infrastructure Security Education for Professionals and Students (CiSE-ProS) project [166]. After participating in a tutorial to learn about the game, players will be introduced to a virtual data center where they can use the available tools to perform safety actions such as safely removing cables and replacing defective RAM cards (Figure 12).

Additionally, the "CybAR" AR game provides an innovative approach to simulating various cyber threats, combining task completion with rewards and feedback to improve cybersecurity defenses among students [39].

This transition from basic to advanced VR/AR experiences in cybersecurity training mirrors the rapid technological advancements in the field. More affordable and user-friendly VR headsets, such as the Oculus Quest, along with the integration of AR and Mixed Reality (XR) technologies, have significantly enhanced the interactivity and engagement of cybersecurity training.

The incorporation of VR/AR in cybersecurity training exemplifies a shift towards experiential learning. By enabling users to interact with virtual threats in a controlled setting, VR/AR has redefined interactive learning in cybersecurity awareness, signaling a trajectory of continuous innovation and enhancement in this domain.

Virtual and Augmented Reality are transforming serious gaming into an immersive and effective tool for cybersecurity awareness [167], [168].

Salazar et al. [114] employed AR to educate high school students about cybersecurity, using a game with physical components to complement learning from an initial presentation. While this method improved self-awareness, it didn't significantly enhance cybersecurity knowledge.

CyberVR [115], utilizing VR, offers an interactive way to learn about IT system threats through mini-games representing various cyber threats. This method was found to be as educational as traditional methods, but more engaging [116].

On mobile platforms, CybAR [117] demonstrates the adaptability of VR/AR for cybersecurity education on accessible devices. This AR game uses a reward-punishment system to teach players about cybersecurity, showcasing the potential of VR/AR in enhancing cybersecurity training.

In summary, the application of VR and AR in cybersecurity training highlights the significant advancements in educational technology, offering a more immersive and engaging learning experience. This evolution is indicative of the broader trend towards incorporating cutting-edge technologies in educational methodologies, especially in fields like cybersecurity where experiential learning is crucial.

## IV. COMPARATIVE ANALYSIS

This section provides an analysis describing the strengths and weaknesses of each of the training methods described throughout this review.

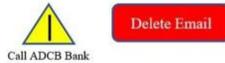

Fig. 9: screenshot of Cyber Shield game showing a phishing simulation with tips to raise phishing awareness [26]

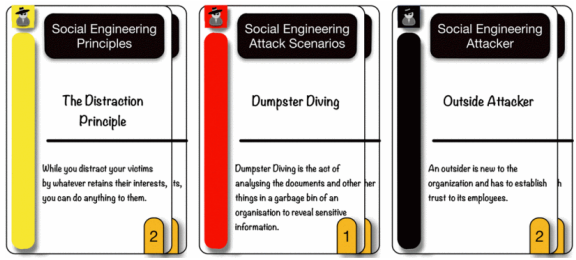

A screenshot of a card game showing SE attack principles [162]

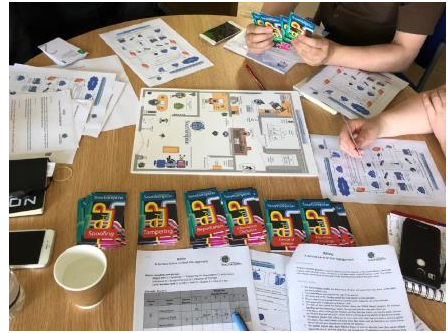

A screenshot of RISKIO tabletop game [74]

A screenshot of Cyber Suraksha card game [163]

Fig. 10: Examples of physical cybersecurity awareness serious games

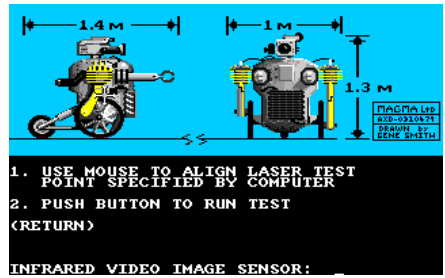

A screenshot of HACKER (1985) game [161]

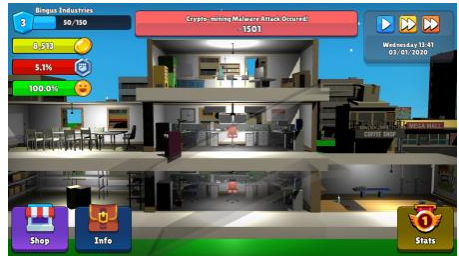

A screenshot of Cyber Security Tycoon game [164]

Fig. 11: Examples of the development of gamification over the past years

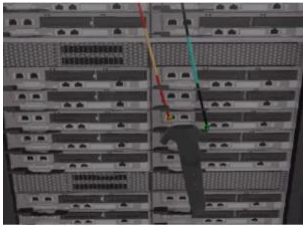

Safely removing cables from the server

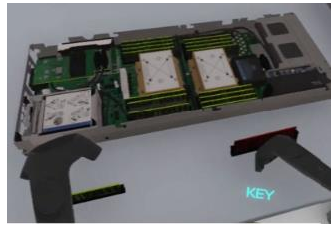

Replacing RAM card

Fig. 12: Screenshots from CISE-PROS VR game [166]

### A. Traditional Training Methods

*1) Passive Awareness Methods:* Passive awareness methods, such as handouts, posters, and electronic communications like e-brochures and weekly e-articles, offer a cost-effective approach to cybersecurity training. These methods benefit from their low-cost nature, primarily due to one-time production expenses, and the potential for cost elimination when utilizing their electronic counterparts. Their scalability is a significant advantage, allowing organizations to disseminate information broadly with minimal effort, reaching a vast audience through digital distribution [169].

However, the accessibility of these methods, while broad due to their simplicity, is often superficial. The static nature of the content can lead to disengagement, as the information may not be sufficiently dynamic to capture the ongoing changes in cybersecurity threats. This lack of personalization and engagement fails to address the specific learning needs or knowledge gaps of individuals, which is crucial in a large, diverse organization [17]. Studies, such as the one conducted by Potgieter, highlight this shortfall, with findings indicating that a significant portion of participants overlooked awareness posters, and electronic newsletters were frequently disregarded or marked as spam [170].

To enhance the effectiveness of passive methods, incorporating interactive elements could combat the inherent engagement challenges. Regular updates are essential to maintain relevance against the backdrop of evolving cyber threats. Strategically placing updated posters and distributing new handouts at regular intervals can help sustain attention and retention. Moreover, integrating discussions of these materials into regular meetings can ensure they are not overlooked and provide opportunities for feedback and improvement.

In essence, while passive awareness methods are highly scalable and accessible, their efficacy is contingent upon dynamic content delivery and periodic refreshment to align with the latest cybersecurity landscapes. By adopting a proactive and interactive approach to these methods, organizations can foster a robust information security culture.

*2) Classroom-Based:* Classroom-based training, encompassing both instructor-led and online modalities, offers a dynamic platform for cybersecurity education. Instructor-led sessions facilitate real-time interaction, fostering an environment conducive to active participation and immediate feedback, which is instrumental in mitigating boredom and disengagement [171]. The personalization of content to match the group's cybersecurity knowledge level is a distinct advantage of face-to-face (FTF) learning, aligning with the schedules of those who prefer structured learning environments [172].

However, the scalability of instructor-led training is inherently limited by physical and logistical constraints, such as venue capacity and instructor availability. While online-based training ameliorates some of these limitations, offering greater reach and flexibility, it requires substantial investment in content creation to ensure engagement [24]. Despite the potential for broader reach, virtual classrooms must strive to replicate the engagement levels of their FTF counterparts to prevent the loss of interest over time [173].

To address these challenges, it is imperative to introduce hands-on projects that complement theoretical learning, thereby enhancing practical understanding and retention. Interactive elements, such as discussions and multimedia tools, should be integrated to maintain enthusiasm and engagement. By tailoring content to the real-time cybersecurity landscape and providing diverse, hands-on experiences, classroom-based training can effectively raise cybersecurity awareness and capabilities [16], [18], [21], [22]. Moreover, regular updates and the inclusion of real-world scenarios will ensure that the training remains relevant and combats the issues of boredom and waning interest [23].

While classroom-based training methods offer a personalized and engaging approach to cybersecurity education, their effectiveness in large organizations hinges on their ability to scale and remain accessible. By leveraging the strengths of both FTF and online modalities, and continually adapting to the evolving cybersecurity threats, these methods can provide a robust foundation for building a strong cybersecurity culture within an organization.

*3) Video-Based:* Video-based training offers a scalable and flexible approach to cybersecurity awareness, allowing learners to engage with content at their convenience. This modality excels in scalability; once produced, videos can be disseminated broadly across various platforms with negligible additional costs, reaching a wide audience with ease [174].

Accessibility is another strength of video-based training. It accommodates self-paced learning, catering to individuals with diverse schedules and learning speeds. However, the one-size-fits-all nature of pre-recorded videos may not address the unique learning styles of all users, potentially affecting the overall efficacy of the training.

To enhance the impact of video-based learning, it is essential to focus on the creation of concise and compelling content. Videos should incorporate interactive elements to engage viewers actively, breaking away from the passivity that often accompanies video viewing. Quizzes and facilitated discussions can be integrated to reinforce learning outcomes and ensure retention. By aligning video content with the learners' objectives and providing opportunities for application through real-world examples, the limitations of individualism and lack of personalization can be mitigated.

While video-based training methods are inherently scalable and accessible, their success in fostering cybersecurity awareness depends on their ability to engage and adapt to the needs of a diverse audience. Through strategic content design and the inclusion of interactive and supplementary materials, these methods can be optimized to deliver effective cybersecurity education.

*4) Cyber Campaigns:* Cyber campaigns represent a dynamic approach to cybersecurity awareness, capable of engaging a broad audience through a variety of interactive methods such as workshops, games, and competitions. These campaigns offer scalability, allowing organizations to reach numerous participants across different locations, particularly when integrating digital and virtual components [175], [176].

The inherent flexibility of cyber campaigns accommodates a spectrum of activities, making them adaptable to a range of preferences and learning styles. This adaptability, however, comes with the caveat of the "one size fits all" model, which may not fully address the nuanced learning needs of every participant. To counteract this, campaigns must be carefully tailored to target the specific cybersecurity threats that are most pertinent to the organization and its workforce.

Effective campaigns should foster an environment of social interaction and personal motivation, encouraging active participation and engagement. This can be achieved by incorporating elements of competition, collaboration, and recognition, which can significantly enhance the learning experience and drive behavioral change [177].

Moreover, the content of cyber campaigns requires regular reviews and updates to ensure that the knowledge imparted remains relevant and is retained over the long term. By aligning campaign activities with the latest cybersecurity challenges and maintaining a pulse on the organization's threat landscape, campaigns can evolve into a powerful tool for instilling a robust cybersecurity culture.

While cyber campaigns are an effective medium for disseminating cybersecurity awareness on a large scale, their success is contingent upon customization, interactive engagement, and ongoing content revitalization. Through strategic planning and execution, cyber campaigns can lead to meaningful improvements in cybersecurity behaviors and knowledge retention.

*B. Technology-Based Training*

*1) Simulation-Based:* Simulation-based training, particularly through the deployment of phishing awareness emails, stands out for its scalability and rapid dissemination capabilities. It enables organizations to efficiently conduct widespread awareness exercises, such as simulated phishing attacks, which can be easily tailored and distributed across various departments and geographic locations [178], [179].

The cost-effectiveness of this method is notable, as it primarily involves the creation and electronic distribution of convincing simulated emails. To ensure ongoing effectiveness and to prevent predictability, it is crucial for organizations to periodically refresh the design and content of these simulations. Moreover, personalization of the phishing scenarios can significantly enhance their impact, as employees tend to be more vulnerable to attacks that resonate with their personal interests or job functions.

However, the method's efficacy may be undermined if the simulations lack sophistication or fail to accurately represent the complexity of real-world phishing attempts. Additionally, accessibility concerns must be addressed, particularly for employees who may not be well-versed in email usage or those working with disparate email systems that could affect the display and interaction with the simulations [151].

To optimize the effectiveness of simulation-based training, it is imperative that the simulated attacks are crafted with a high degree of realism, utilizing social engineering techniques to mirror genuine phishing attempts. The study by Abdullah et al. [151] underscores this point, demonstrating a significant increase in employee engagement with the simulated emails when they were meticulously designed to reflect legitimate correspondence (Figure 7).

While simulation-based training is a powerful tool for enhancing information security awareness, its success is contingent upon the authenticity of the simulations, the adaptability to the target audience, and the ability to evolve with the changing landscape of cybersecurity threats.

*2) App-Based:* App-based training, akin to the game-based approach, offers an engaging alternative to traditional cybersecurity education methods. It provides the convenience of self-paced learning, allowing users to engage with the material on their own terms and schedules. The versatility of these applications, which are compatible across a range of devices including mobiles and web platforms, enhances their ubiquity and flexibility, making them a practical choice for widespread cybersecurity education [40].

The ease of use associated with mobile learning (M-Learning) tools is one of their strongest attributes, often requiring minimal technical expertise, thus broadening their appeal across diverse user groups. These applications typically offer a structured learning experience that extends beyond theoretical knowledge, integrating practical scenarios that users are likely to encounter [95], [180].

However, the uniformity of content in app-based training could be a limiting factor, as it may not adequately address the varied needs of different user groups within an organization. The assumption that all users will benefit from the same training materials overlooks the nuances of individual technical backgrounds and learning preferences.

Moreover, accessibility remains a potential barrier, particularly for individuals who may not possess compatible devices or who are less inclined to utilize mobile applications for educational purposes. This highlights the need for a more inclusive approach that considers the technological readiness and preferences of all employees.

To maximize the effectiveness of app-based cybersecurity training, it is essential to maintain a dynamic and updated content strategy. This includes regular updates to the training material to reflect the latest cybersecurity threats and enhanced graphics to sustain user engagement and prevent monotony. Additionally, ensuring cross-platform functionality can significantly improve accessibility, allowing users to interact with the training material on their preferred devices, be it smartphones, PCs, or through web interfaces [87].

### C. Innovative Training Methods

*1) Game-Based Training:* Serious gaming, recognized for its engaging and interactive nature, offers a dynamic approach to cybersecurity education. It is particularly noted for its potential to foster long-term knowledge retention [181]. The core advantage of game-based methods lies in their ability to stimulate active learning and provide immersive experiences through hands-on interaction. By leveraging elements of competition, reward systems, and interactive storytelling, games can significantly boost motivation and engagement among participants [182], [183].

Cybersecurity games often feature progressive difficulty levels, allowing users to incrementally build their awareness and skills. The inclusion of both multiplayer and single-player modes caters to social interaction and individual progression, respectively. Real-time feedback mechanisms within these games serve to guide players, correcting actions and suggesting improved strategies [184]. Furthermore, the flexibility of serious games to function across various platforms, including mobile devices and web interfaces, enhances their accessibility to a diverse user base.

However, scalability and content relevance pose significant challenges in game-based training. Tailoring games to cater to the specific roles and skill levels within a large organization can be a complex and costly endeavor. The development, maintenance, and regular updating of these games require substantial investment, which may not always be justifiable, especially when considering the diverse needs of a heterogeneous workforce [185].

Moreover, while personalization is a sought-after feature in serious games, achieving it at an individual level remains a formidable task. Games like "PeriHack" demonstrate the potential for variety in cybersecurity challenges, yet the extent to which such games can be personalized to address every user's unique learning needs is limited [186].

Demographic considerations also come into play; not all employees may resonate with a gaming approach to learning. Older employees or those from regions with limited internet access may find traditional methods more conducive to their learning style.

To enhance the efficacy of game-based cybersecurity training, developers must focus on maintaining the quality and relevance of the game content. This involves not only the aesthetic appeal, such as graphics, but also the educational substance, such as the number of levels and the diversity of cybersecurity scenarios presented. Continuous updates and the inclusion of feedback mechanisms are crucial to address the evolving landscape of cyber threats and to keep the players informed of their progress [187].

While game-based training offers an innovative approach to cybersecurity awareness, it requires careful consideration of scalability, cost, accessibility, and demographic diversity to ensure its effective implementation within an organization.

*2) Virtual Reality/Augmented Reality:* Virtual and Augmented Reality (VR/AR) technologies are at the forefront of immersive learning, offering environments that closely mimic real-life scenarios. Studies such as those by [188] and [189] have demonstrated the superior efficacy of VR in knowledge acquisition and retention when compared to more traditional methods. For instance, "Hack the Room," an AR-based game, has shown promising results in teaching the principles of ethical hacking, significantly enhancing knowledge retention among its users [190].

However, despite these advantages, VR and AR are the least scalable of the cyber training methods due to the high costs associated with their implementation. The need for specialized equipment such as VR headsets, 360 cameras, and high-performance computing resources presents a considerable barrier [191]. Moreover, the accessibility of VR/AR training is limited by the availability of this equipment to all employees, as well as additional considerations such as physical space requirements and potential health implications like motion sickness.

The quality of the VR/AR experience is paramount; it hinges on the sophistication of the virtual environment and the interactivity it affords. While VR/AR can significantly enhance engagement levels compared to other interactive methods like serious gaming, the initial and ongoing investment in technology and content development is substantial. This method also demands considerable time for both the training of personnel and the maintenance of the equipment.

To optimize the use of VR/AR in cybersecurity training, organizations must weigh the benefits against the costs and accessibility challenges. While the immersive nature of VR/AR can lead to high retention rates, the practicality of widespread implementation must be carefully considered. Organizations should evaluate whether the enhanced learning outcomes justify the significant investment in specialized hardware and the potential limitations on accessibility. As with any educational tool, the ultimate goal is to provide effective learning experiences that are both engaging and accessible to the widest possible audience.

In the comparative analysis of training delivery methods, see Table V and Table IV, it is evident that there is no one-size-fits-all solution. Each method presents a unique blend of strengths and weaknesses across various metrics such as cost, reachability, time effectiveness, personalization, engagement, ease of implementation, knowledge retention, scalability, and accessibility.

Physical methods like conventional and classroom settings offer high engagement and personalization but often at a higher cost and lower scalability. Campaigns, while cost-effective and with a broader reach, may lack personalization and knowledge retention.

Virtual/digital methods offer distinct advantages in reachability and scalability, particularly with conventional online methods and videos, which also tend to be cost-effective. However, these methods can fall short in engagement and personalization unless specifically designed to address these areas, as seen in gamification and VR/AR methods.

The choice of method should be guided by a strategic alignment with the organization's learning objectives, audience characteristics, budgetary constraints, and desired outcomes. High scalability and reachability of digital methods may be favored by organizations with a widespread workforce, while those prioritizing deep skill acquisition might lean towards high personalization and engagement offered by classroom or VR/AR methods, despite the higher cost.

Ultimately, a blended approach that leverages the strengths of multiple methods may often be the most effective strategy. This approach allows organizations to tailor their training programs to the specific needs of their workforce, ensuring that each individual receives the most appropriate form of training to maximize learning outcomes and retention. The key is to balance these factors to achieve the desired educational impact while remaining within the operational constraints.

## V. Emerging Trends

This section discusses the emerging trends that are shaping the future of cybersecurity training, and what implementation challenges exist.

### A. Artificial Intelligence (AI)

Artificial Intelligence (AI) and its subsets, Machine Learning (ML) and Deep Learning (DL), have become integral to our daily lives, offering automation of tasks, complex problem-solving, and adaptability with minimal human intervention. In cybersecurity, AI's capabilities are particularly transformative, enhancing traditional defense methods, which often rely heavily on human intervention. AI facilitates automatic risk identification amidst the deluge of cyber threats, defends against automated bot attacks, predicts future incursions, and adapts to counter novel threats that traditional antivirus software may not recognize [192]. While conventional antivirus solutions are adept at defending against known threats through pattern and signature detection, they falter against advanced threats like metamorphic malware, which can alter its appearance to evade detection [193], [194]. Machine and Deep Learning algorithms, such as Neural Networks, Support Vector Machines, and K-Nearest Neighbours, offer a more robust approach by identifying anomalous behavior within networks (anomaly-based detection) and providing predictive analyses of potential future attacks, rather than merely comparing against known signatures [195]–[198].

Beyond serving as a countermeasure against cyberattacks, AI has been widely adopted as a preventive measure to enhance cyber awareness and analyze human behavior. Tools like ViCyber, an intelligent cloud-based system, aim to tailor cybersecurity curricula to users' knowledge and interests using visual mapping and nearest neighbor classification algorithms to evaluate the suitability of selected curricula [199], [200]. Similarly, the "Sifu" platform utilizes AI to assess code for vulnerabilities, employing security testing tools like Static Application Security Testing (SAST) and Dynamic Application Security Testing (DAST), thereby providing personalized feedback to users [201].

The application of Reinforcement Learning (RL) and Deep Reinforcement Learning (DRL) introduces a higher degree of adaptability and personalization by learning from environmental feedback and reward signals. These algorithms are adept at real-time cyber-attack detection, defense against DoS and DDoS attacks, phishing and social engineering attack recognition, and enhancing user cybersecurity awareness [202]–[206]. However, while the benefits of AI in cybersecurity are substantial, it is imperative to address the ethical considerations and the potential for false positives in behavior-based

TABLE IV: Comparative analysis of the delivery methods when conducted physically

| Method \ Metrics | Cost | Reachability | Time Effectiveness | Personalization | Engagement | Ease of Implementation | Knowledge Retention | Scalability | Accessibility |
|---|---|---|---|---|---|---|---|---|---|
| Conventional | Medium | Medium | High | Low | Low | High | Medium | High | Medium |
| Classroom | High | Low | Medium | High | High | Medium | High | Low | High |
| Campaigns | Low | High | Low | Low | Medium | Low | Low | Medium | Medium |

TABLE V: Comparative analysis of the delivery methods when conducted virtually/digitally

| Method \ Metrics | Cost | Reachability | Time Effectiveness | Personalization | Engagement | Ease of Implementation | Knowledge Retention | Scalability | Accessibility |
|---|---|---|---|---|---|---|---|---|---|
| Conventional | Low | High | Low | Low | Low | High | Low | High | High |
| Classroom | Medium | Low | High | High | Low | Medium | Medium | Medium | High |
| Gamification | Medium | High | Medium | Medium | High | Low | High | Low | Medium |
| Simulation | High | Medium | High | High | Low | Medium | Medium | High | Low |
| Application | Medium | High | Low | Low | Low | Low | Low | Medium | Medium |
| Videos | Low | High | High | Low | Low | High | Low | High | High |
| VR/AR | High | Medium | Medium | High | High | Low | High | Low | Low |

AI systems. The predictive nature of AI, while powerful, can lead to misclassifications, inadvertently flagging benign behavior as malicious. This necessitates the implementation of robust validation mechanisms to minimize false positives and ensure ethical use of AI, particularly when monitoring employee behavior to safeguard against insider threats [207]–[210].

The challenges of data bias, imbalance, and overfitting are significant when benign operations outnumber malicious ones. To address this, data augmentation techniques, such as those provided by Generative Adversarial Networks (GANs), are employed. GANs, with their dual components of Generators and Discriminators, are instrumental in creating synthetic data to train models, enhancing their ability to detect real-world anomalies [211]–[213].

While RL and DRL have demonstrated efficacy in cyber defense, their limitations should not be overlooked. RL struggles with complex, high-volume data and requires pre-existing datasets for learning, necessitating data augmentation to prevent imbalance. Conversely, DRL can handle large datasets more effectively without the need for pre-existing data. However, it may incur higher maintenance costs and risk overfitting due to its intensive data processing capabilities. To mitigate these drawbacks, future research should focus on developing cost-effective DRL models that are resilient to overfitting and capable of handling the intricacies of cybersecurity data without compromising performance.

*B. Extended Reality (XR)*

Extended Reality (XR) encompasses the spectrum of Virtual Reality (VR), Augmented Reality (AR), and Mixed Reality (MR), technologies that blend human-machine interaction within virtual environments. As delineated in previous sections, VR and AR have shown promise in elevating cybersecurity awareness by simulating cyber-attack scenarios, thereby offering a multi-sensory digital experience that surpasses traditional learning methods in engagement and retention. MR, in particular, transcends the limitations of VR and AR by not only overlaying digital objects onto the real world but also allowing users to interact with these objects within a fully-immersive digital environment, thus creating a more profound educational impact [214], [215].

The prototypes like "CS:NO" and CyVR-T, as well as the PlantVR and PlantAR, exemplify the potential of XR to visualize and interact with cyber threats in a controlled, educational setting, enabling users to construct and navigate through their own cyber-attack scenarios [216], [217]. These applications represent a leap forward from traditional cybersecurity training methods, which often struggle with engagement and flexibility. XR bridges this gap by combining the best of online and face-to-face (FTF) classes, offering a virtual space for interactive learning that can be both flexible and engaging [217]–[219].

However, the discussion would be incomplete without addressing the accessibility and scalability of XR technology. While XR offers an innovative approach to cybersecurity training, its widespread adoption is contingent upon the technology being both accessible and scalable. Accessibility pertains to the ease with which users can obtain and utilize XR technology, which is currently hindered by the cost of equipment and the need for technical expertise. Scalability, on the other hand, involves the technology's capacity to be expanded and adapted to large numbers of users across diverse settings.

The current state of XR technology presents challenges in both areas. High-quality XR experiences often require sophisticated hardware and software, which can be cost-prohibitive for many institutions. Moreover, the development of XR content is resource-intensive, requiring specialized skills that may not be readily available. To address these challenges, research and development efforts must focus on creating cost-effective XR solutions and simplifying content creation [220]. This could involve the development of more affordable hardware, open-source software platforms, and user-friendly tools for educators to create their own XR content.

Furthermore, serious gaming and educational applications utilizing MR have demonstrated the potential for interactive and tangible learning experiences. Games that simulate firewall security, encryption-decryption processes, and phishing detection not only engage students but also provide hands-on experience with cybersecurity concepts [214], [221]–[223]. To scale these experiences for widespread use, the development of standardized XR platforms that can be customized for different educational contexts is essential. These platforms must be designed with inclusivity in mind, ensuring that they are accessible to users with varying levels of technical proficiency and across different types of devices.

While XR technology holds significant promise for cybersecurity training, its full potential will only be realized through concerted efforts to enhance its accessibility and scalability. By addressing these challenges, XR can become a cornerstone of cybersecurity education, providing immersive and interactive experiences that prepare users to face the evolving landscape of cyber threats.

When comparing these emerging trends to the methods discussed in Section IV, see Table VI, one can discern a clear trajectory towards increased effectiveness and efficiency. AI and ML technologies can automate and optimize the detection and response to cyber threats, thereby enhancing the efficiency of cybersecurity training. They offer a level of personalization that is difficult to achieve in traditional settings, addressing the individual's unique learning curve and potentially increasing the effectiveness of the training.

XR technologies, while currently less scalable due to high costs and technological barriers, offer effectiveness in knowledge retention that traditional methods struggle to match. The immersive nature of XR can simulate real-world scenarios, providing hands-on experience in a controlled environment, which is crucial for preparing individuals to respond to actual cyber threats.

In terms of efficiency, while the initial setup for XR might be resource-intensive, the long-term benefits of enhanced learning experiences could justify the investment. The scalability challenge of XR could be mitigated as the technology becomes more mainstream and cost-effective, potentially offering a more efficient way to train large numbers of individuals in diverse locations.

While traditional methods provide a solid foundation for cybersecurity training, the integration of AI, ML, and XR technologies represents a significant advancement in both the effectiveness and efficiency of such programs. These emerging trends, with their ability to provide personalized, engaging, and real-time training experiences, are well-positioned to become the cornerstone of cybersecurity education and defense strategies. As these technologies mature and become more accessible, they are likely to set new benchmarks for cybersecurity training, surpassing traditional methods in their ability to prepare individuals and organizations to face the cyber threats of tomorrow.

## VI. CHALLENGES AND FUTURE DIRECTIONS

Employing the different IS training methods described throughout this paper has several challenges that should be considered. These difficulties are listed next and future directions for research are then provided.

---

[1]Cost-Efficiency refers to the long-term value relative to the effectiveness of the training.

[2]Scalability for emerging trends is expected to improve as the cost and accessibility of technology improve.

[3]Learning retention for emerging trends is suggested to be higher based on the immersive and interactive nature of the training, as indicated by preliminary studies.

[4]Maintenance considerations include the need for regular updates and the technical expertise required for system upkeep.

### A. Resource Allocation

A perennial challenge is the underestimation of cybersecurity training's importance, often perceived as a non-urgent expense [224]. This misperception leads to inadequate funding and time allocation, which are critical for the adoption of effective cybersecurity measures [225]. The scarcity of resources extends to the recruitment of skilled personnel and the acquisition of necessary software and hardware. Small and Medium Enterprises (SMEs) in particular face significant barriers, with budget constraints and competing business priorities impeding the implementation of robust cybersecurity plans [226]. Future research should explore cost-effective cybersecurity training models, particularly for SMEs, and develop frameworks for conducting cost-benefit analyses that quantify the long-term value of cybersecurity investments [25].

### B. Adapting to Evolving Threats

Cyber threats are in a state of constant evolution, with ransomware and sophisticated phishing attacks presenting ongoing challenges [227]. The dynamic nature of these threats necessitates continuous updates to training content, a process that can be resource-intensive. Research should focus on developing adaptive training modules that can quickly integrate updates on new threats, leveraging AI to automate content generation and reduce the overhead associated with maintaining up-to-date training materials [228].

### C. Engagement and Motivation

Low participation rates in IS training are often attributed to a lack of motivation and the perception of training as a checkbox exercise rather than an ongoing process [17]. To combat this, organizations must innovate training programs that are both engaging and rewarding. Incorporating real-world attack simulations can enhance relevance and interest [81]. Future research should investigate the impact of gamification and interactive content on participation rates and identify motivational drivers that encourage proactive engagement with cybersecurity training [229].

### D. Balancing Technological and Training Investments

Organizations frequently prioritize direct business growth initiatives, such as the acquisition of new software or hardware, over cybersecurity training [230]. This can create vulnerabilities, as sophisticated cyber threats often circumvent traditional security tools. Future research should examine strategies for achieving a balance between technological investments and cybersecurity training, emphasizing the importance of human factors in maintaining organizational security.

### E. Developing Effective Metrics

The absence of robust metrics to measure the impact of cybersecurity training hinders the ability to evaluate and improve training methods [231]. Research is needed to establish comprehensive metrics that can assess behavioral changes and the return on investment in cybersecurity training. These metrics should inform continuous improvement processes and support the justification of training budgets.

TABLE VI: Comparative Analysis of Traditional Methods and Emerging Trends in Cybersecurity Training

| Metrics | Traditional Methods | Emerging Trends (AI, ML, XR) |
|---|---|---|
| Engagement | Low to Moderate | High |
| Personalization | Low | High |
| Adaptability | Low | High |
| Predictive Capability | None to Low | High |
| Immersiveness | None to Low | High |
| Cost-Efficiency[1] | Moderate to High | High (decreasing with tech advances) |
| Scalability[2] | Moderate | Variable (improving with tech accessibility) |
| Learning Retention[3] | Moderate | High (based on preliminary studies) |
| Real-time Application | Low | High |
| Maintenance[4] | Low to Moderate | Moderate to High (depending on system complexity) |

*F. Inclusive and Multilingual Training*

The reliance on English-language sources limits the global applicability of cybersecurity training. Future research should extend to multilingual studies, incorporating diverse linguistic and cultural contexts to ensure inclusivity and global relevance in cybersecurity training methodologies.

## VII. CONCLUSION

This comprehensive survey has dissected the multifaceted landscape of cybersecurity awareness training, offering a critical lens through which the evolution, effectiveness, and future trajectory of various methodologies are examined. Our systematic approach, underpinned by a robust literature review and enriched by extensive online searches and seasoned personal observations, has yielded a nuanced understanding of the field's current state and emerging trends.

We have traversed the spectrum from traditional to cutting-edge training methods, revealing that while conventional strategies lay the groundwork for awareness, innovative approaches leveraging artificial intelligence, virtual reality, and gamification are poised to redefine engagement and effectiveness in cybersecurity training. Our comparative analysis has illuminated the strengths and weaknesses inherent in each method, providing a strategic vantage point for practitioners and scholars alike.

Emerging trends such as AI, VR, and AR have been identified as harbingers of a new era in cybersecurity training, promising to enhance adaptability and immersion. However, these advancements are not without their implementation challenges, which we have addressed with practical solutions drawn from real-world applications.

Our contribution to the field is distinct. We have not only cataloged and evaluated current methodologies but also set forth a vision for the future, anticipating developments and directing scholarly inquiry toward uncharted territories in cybersecurity awareness training. This paper has filled gaps left by previous literature, offering a comprehensive, comparative, and critical analysis that stands as a beacon for future research.

In essence, this survey serves as a cornerstone for ongoing and future discourse in cybersecurity training, providing a scaffold upon which resilient cyber defenses can be built, not just through technology, but through the empowerment of human vigilance and adaptability in the face of ever-evolving cyber threats.

REFERENCES

[1] S. Kemp, "Digital 2023: Global overview report.," 2023. Accessed on Mar 25, 2023.

[2] Q. Hou, M. Han, and Z. Cai, "Survey on data analysis in social media: A practical application aspect," *Big Data Mining and Analytics*, vol. 3, no. 4, pp. 259–279, 2020.

[3] A. E. E. Sobaih, A. M. Hasanein, and A. E. Abu Elnasr, "Responses to covid-19 in higher education: Social media usage for sustaining formal academic communication in developing countries," *Sustainability*, vol. 12, no. 16, p. 6520, 2020.

[4] Z. Mu, X. Liu, and K. Li, "Optimizing operating parameters of a dual e-commerce-retail sales channel in a closed-loop supply chain," *IEEE Access*, vol. 8, pp. 180352–180369, 2020.

[5] O. Sviatun, O. Goncharuk, C. Roman, O. Kuzmenko, and I. V. Kozych, "Combating cybercrime: economic and legal aspects," *WSEAS Transactions on Business and Economics*, vol. 18, pp. 751–762, 2021.

[6] D. J. Borkovich and R. J. Skovira, "Working from home: Cybersecurity in the age of covid-19.," *Issues in Information Systems*, vol. 21, no. 4, 2020.

[7] A. Georgiadou, S. Mouzakitis, and D. Askounis, "Working from home during covid-19 crisis: a cyber security culture assessment survey," *Security Journal*, vol. 35, no. 2, pp. 486–505, 2022.

[8] F. Salahdine and N. Kaabouch, "Social engineering attacks: A survey," *Future Internet*, vol. 11, no. 4, p. 89, 2019.

[9] H. Aldawood and G. Skinner, "Analysis and findings of social engineering industry experts explorative interviews: perspectives on measures, tools, and solutions," *IEEE Access*, vol. 8, pp. 67321–67329, 2020.

[10] S. Venkatesha, K. R. Reddy, and B. Chandavarkar, "Social engineering attacks during the covid-19 pandemic," *SN computer science*, vol. 2, pp. 1–9, 2021.

[11] M. F. Ansari, P. K. Sharma, and B. Dash, "Prevention of phishing attacks using ai-based cybersecurity awareness training," *Prevention*, 2022.

[12] K. Shaukat, S. Luo, V. Varadharajan, I. A. Hameed, S. Chen, D. Liu, and J. Li, "Performance comparison and current challenges of using machine learning techniques in cybersecurity," *Energies*, vol. 13, no. 10, p. 2509, 2020.

[13] H. Aldawood and G. Skinner, "Reviewing cyber security social engineering training and awareness programs—pitfalls and ongoing issues," *Future Internet*, vol. 11, no. 3, p. 73, 2019.

[14] H. Aldawood and G. Skinner, "Contemporary cyber security social engineering solutions, measures, policies, tools and applications: A critical appraisal," *International Journal of Security (IJS)*, vol. 10, no. 1, p. 1, 2019.

[15] M. Alshaikh, "Developing cybersecurity culture to influence employee behavior: A practice perspective," *Computers & Security*, vol. 98, p. 102003, 2020.

[16] E. Kweon, H. Lee, S. Chai, and K. Yoo, "The utility of information security training and education on cybersecurity incidents: an empirical evidence," *Information Systems Frontiers*, vol. 23, pp. 361–373, 2021.

[17] W. He and Z. Zhang, "Enterprise cybersecurity training and awareness programs: Recommendations for success," *Journal of Organizational Computing and Electronic Commerce*, vol. 29, no. 4, pp. 249–257, 2019.